\documentclass[11pt,twocolumn]{article}

\usepackage{graphicx}
\usepackage{amsmath,amssymb,textcomp}
\usepackage{subfigure}
\usepackage{cite}
\usepackage[pdfstartview=FitH,colorlinks]{hyperref}
\usepackage{url,array}

\newcommand{\beq}{\begin{eqnarray}}
\newcommand{\eeq}{\end{eqnarray}}

\newcommand{\cm}{\text{cm}}
\newcommand{\pc}{\text{pc}}
\newcommand{\s}{\text{sec}}

\begin{document}

\title{Gamma-ray evidences of the dark matter clumps}

\author{K.~Belotsky\thanks{k-belotsky@yandex.ru} \and A.~Kirillov\thanks{kirillov-aa@yandex.ru} \and M.~Khlopov\thanks{khlopov@apc.univ-paris7.fr}}

\date{}

\twocolumn[
	\begin{@twocolumnfalse}
	\maketitle
	\abstract{
		We discuss the possibility of identification of point-like gamma-ray sources (PGS) with small scale dark matter (DM) clumps in our Galaxy. Gamma-rays are supposed to originate from annihilation of DM particles in the clumps, where annihilation rate is supposed to be enhanced, besides higher density, due to smaller relative velocities $v$ of DM particles. We parameterized the annihilation cross section $\sigma_\text{ann}(v)$ in the form of an arbitrary power law dependence on the relative velocity $v$ with/without factor of Sommerfeld-Gamow-Sakharov, implying existence of a new Coulomb-like interaction. Adopting different parameters of cross section and clump, satisfying condition $\Omega\lesssim 0.2$ on density of DM particles of question, they are constrained from comparison with Fermi/LAT data on unidentified PGS as well as on diffuse $\gamma$-radiation; results are applied to concrete DM candidates.
		Such analysis is found to be sensitive enough to existing uncertainty in the density profiles of DM in the clump what can provide a tool for their test.
		Also we discuss possibilities when gamma-radiating clump changes visibly its position on celestial sphere and it is seen as a spatially extended gamma-source (EGS), what can be probed in future experiments like Gamma-400.
	}
	\smallskip

	{\bf Keywords:} dark matter, nonidentified gamma-ray sources, gamma-rays
	\bigskip

  \end{@twocolumnfalse}
]
{
  \renewcommand{\thefootnote}%
    {\fnsymbol{footnote}}
  \footnotetext[1]{k-belotsky@yandex.ru}
  \footnotetext[2]{kirillov-aa@yandex.ru}
  \footnotetext[3]{khlopov@apc.univ-paris7.fr}
}


\section{Introduction}

From the first articles revealing the indirect effects of the cold dark matter (CDM) in the 
form of heavy neutral leptons \cite{1977PhRvL..39..165L, 1977PhRvL..39..168D, 1978ApJ...221..327D, 1978ApJ...223.1015G, 1978ApJ...223.1032S, 1980YadPhys...31..1286, 1984PhRvL..53..624S, 1994PAN....57..425K} or supersymmetric particles \cite{1984PhRvL..53..624S, 1985PhRvL..55.2622S, 1988ApJ...325...16R}, such indirect effects had been the subject of intensive studies in the cosmic ray (CR) data.
It was shown that the DM particles could form the hierarchic structures over a wide range of scales and masses (from small scale clumps to large scale structures) \cite{1977ApJ...218..592G, 1978MNRAS.183..341W, 1982ApJ...263L...1P, 1982Natur.299...37B, 1984Natur.311..517B, 1995SvPhU..38..687G}. The annihilation of DM particles within these structures can give the cosmic ray signals \cite{1978ApJ...223.1015G, 1978ApJ...223.1032S, 1980YadPhys...31..1286, 1984PhRvL..53..624S, 1988PhLB..214..403E, 1991PhRvD..43.1774K, 1994PAN....57..425K, 1992PhLB..294..221B, 1995PhLA..208..276G, 2003PhRvD..68j3003B, 2005Natur.433..389D}.
In the clumps the annihilation rate should be enhanced due to higher density and possibly due to amplification of annihilation cross section at small relative velocities of DM particles, which is especially small in the lightest clumps, which are likely to be the most abundant.
The mentioned factors can lead to that the clumps, located in a neighborhood of Solar system, are manifested as discrete (basically as point-like) gamma-ray sources \cite{1992PhLB..294..221B, 1997SvPhU..40..869G, 2003PhRvD..68j3003B, 2009baksan.conf..21, 2013YadPhys...76...286}.

In this paper we continue the previous study of given effect \cite{2013YadPhys...76...286} and, mainly, make more accurate calculations, take into account diffuse $\gamma$-radiation, consider possibility of observation of spatially extended sources in the light of future experiments.

\section{DM annihilation in the CDM clumps}
\label{sec:flux}

Predictions of density profile inside the clumps suffer with some uncertainties \cite{2003IJMPD..12.1157T, 2004ApJ...601...47A, 2006AJ....132.2685M, 2011ASL.....4..297D}. For the most of calculations in this paper we use profile obtained in \cite{1992PhLB..294..221B, 1997SvPhU..40..869G, 2003PhRvD..68j3003B}, which gives rather minimal prediction for $\gamma$-flux from the clump. Comparison with other profile models is given below in terms of the numbers of predicted PGS.
The chosen profile (indicated further as BGZ) was taken in the form
\beq
    \rho(r) =\rho_1
    \begin{cases}
        1, &r\le r_0 \\
        \left(r_0/r\right)^{\eta},&r_{0}<r\le R
    \end{cases}
    \label{eq:ProfBDE}
\eeq
where $\eta=1.8$, $r_0=0.05 R$, $R\approx 10^{18}(M/M_{\odot})^{1/3}$~cm with $M$ being the clump mass \cite{1992PhLB..294..221B,2003PhRvD..68j3003B}.

It is supposed that only minor component of clumps can survive until the present time \cite{2003PhRvD..68j3003B}.
For our estimation $\xi = 0.002$ is taken as the clumps fraction from the full density of DM in Galaxy.
Here we study $\gamma$-radiation effect for only minimal clump mass, formally assuming that all DM clumps are with this mass. In fact, they are predicted to be most abundant \cite{1992PhLB..294..221B, 2003PhRvD..68j3003B, 2010ApJ...723L.195I}, however effect of clumps (sub-halos) from a high-mass tail of mass distribution can be also noticeable \cite{2012ApJ...747..121A}.

A smallness of relative velocity $v$ of DM particles concentrated in the clumps may strongly affect to annihilation rate if the corresponding cross section $\sigma_\text{ann}$ depends from $v$ \cite{2010PhRvD..82i5007C, 2011PhRvD..83l3511H}. We choose the cross section at the parameterized form given below, to cover a wide class of models of DM particles:
\beq
	\sigma_{\text{ann}}=\frac{\sigma_0}{v^{\beta}} C(v,\alpha).
	\label{sigma_common}
\eeq
Here $\beta$ is a free parameter taking (for generality) the continuous range of values (as usual $\beta=1$ means the $s$-wave amplitude only, $\beta=-1$ means the amplitude from $s$- and $p$-waves), $\sigma_0$ at given $\beta$ is determined by cosmological density of the particles $\Omega$. The factor $C(v,\alpha)$ takes explicitly into account a possible Coulomb-like interaction (we will refer to it as ``y-interaction'') of DM particles, which leads to a Sommerfeld-Gamow-Sakharov enhancement \cite{Sommerfeld, 1928ZPhy...51..204G, 1931AnP...403..257S, 1948ZETF...18..631} and has the form:
\beq
    C\left(v,\alpha\right) = \frac{2\pi\alpha/v}{1-\exp\left(-2\pi\alpha/v\right)}.
    \label{eq:coulomb}
\eeq
Here $\alpha$ is the fine structure constant of additional interaction. Such interaction may influence (decrease) relic density a little, but may significantly enhance the annihilation in the present Universe where the particle velocities are small \cite{2005GrCo...11...27B, 2010PhLB..687..275D, 2010PhRvD..81h3502Z, 2010PhRvD..82h3525F, 2011arXiv1107.3546S, 2011PhRvD..83l3513Z}. Annihilation effects become noticeable even for a subdominant component with $\Omega\ll\Omega_{\text{CDM}}$ as it takes place in case of heavy neutrinos \cite{2008PAN....71..147B} to be shown below.

We suppose that an active (annihilating) component of DM may be both dominant and subdominant, i.e. $\Omega\le\Omega_{\text{CDM}}$ with $\Omega_{\text{CDM}}\sim 0.2$ being the total relative density of cold dark matter in Universe. In estimation of cosmological density $\Omega$ we follow to the standard approach \cite{1981RvMP...53....1D, 1986PhRvD..33.1585S}.
It is worth to note that the given scheme does not take into account possibility of binding pairs of considered particle-antiparticle due to y-interaction. The rate of such recombination may exceed expansion rate due to high recombination cross section, and this process becomes crucial, if DM particles decoupled from ambient plasma, leading inevitably to annihilation. It may strongly suppress abundance of these DM particles \cite{2005GrCo...11...27B}.

\section{DM clumps as gamma-ray sources}
\label{sec:flux}

In our estimations it is assumed that DM particle of question is of Dirac type\footnote{The case of Majorana particles has no differences of principle, except for the case of Coulomb-like interaction which is excluded for Majorana particles.} with mass $m\sim 100$~GeV. If annihilating DM particles do not constitute all DM ($\Omega<\Omega_\text{CDM}$) then their contribution to density of clumps is assumed to be proportional to $\Omega/\Omega_\text{CDM}$ (eq.~\eqref{eq:concentration}). We do not specify annihilation channel of photon production, assuming that their averaged multiplicity for energy $E_{\gamma}>100$~MeV is $N_\gamma = 10$. It is quite typical value for high energy processes at respective energy release.

\begin{table*}[t]
\renewcommand{\arraystretch}{2}
    \centering
    \begin{tabular}{cccccc}
        \hline\hline
        Model & \parbox[c]{2cm}{Cored Isothermal} \cite{1991MNRAS.249..523B} & Burkert \cite{1995ApJ...447L..25B} & NFW \cite{1997ApJ...490..493N} &  Einasto \cite{2005frco.conf..241E}  & Moore \cite{1999MNRAS.310.1147M} \\[2mm]
        \hline
        $N/N_\text{BGZ}$ & 1.2 & 6.0$\times10^2$ & $2.1\times10^3$ & $3.3\times10^3$ & $1.7\times10^4$ \\[2mm]
        \hline\hline
    \end{tabular}
	\caption{The number of PGS as predicted for some density profiles in the unites of that for BGZ profile.}\label{tab}
\end{table*}

The photon flux at distance $l$ from the clump center is given by
\beq
	F = \frac{P}{4\pi l^2}= \frac{N_\gamma}{4\pi l^2} \int\limits_{V} \left\langle \sigma_\text{ann}v \right\rangle n\bar{n}\ dV,
	\label{eq:flux}
\eeq
where the particles/antiparticles number density is
\beq
    n = \bar{n} = \frac{1}{2} \frac{\rho(r)}{m}\cdot \frac{\Omega}{\Omega_{\text{CDM}}}.
    \label{eq:concentration}
\eeq
Note that the fraction of subdominant DM particles should be suppressed in the clumps of mass $M<M_{\rm min}$, if they are, where $M_{\rm min}$ is the minimal mass which could be formed by considered DM particles if they prevailed in density. In our study we do not take into account this.

The value $\left\langle\sigma_\text{ann}v\right\rangle$ is determined taking into account velocity distribution of DM particles inside the clump which is assumed to be Maxwellian one with ``virial'' temperature $T_\text{vir}=GMm/2R$.

$\gamma$-Radiation may be registered by LAT, if $E_{\gamma}>100$~MeV \cite{2009ApJ...697.1071A} and their flux exceeds the point source sensitivity $F_{\text{min}}\approx 3\cdot 10^{-9}$~$\cm^{-2}\s^{-1}$. The value $F_{\text{min}}$ allows to calculate the maximal distance $l_\text{max}$ at which the clump can be registered as $\gamma$-source. It gives for BGZ $l_\text{max}\sim 10^{-3}$~pc for $\beta=1.5$, $\sigma_0=10^{-35}$~cm$^2$ and $10^{-10}M_\odot$ without y-interaction and $l_\text{max} \sim 10$~pc with one and the same parameters, and for the Moore model $l_\text{max}\sim 10^{-1}$~pc and $l_\text{max}\sim 10^{2}$~pc respectively. All the obtained $l_\text{max}\ll$~Galactic size, what justifies assumption that clump number density $n_\text{cl}\approx\text{const}$ and corresponds to the local one.

The number of clumps which may be detected by LAT is
\beq
    N_{\text{cl}} =
    n_{\text{cl}}\cdot \frac{4}{3} \pi l_{\text{max}}^3,
    \label{eq:N}
\eeq
where
\begin{equation}
	n_{\text{cl}}=\frac{\xi \rho_{\text{loc}}}{M} \approx 1.6\cdot 10^{-5} \frac{M_\odot}{M}\ \pc^{-3}
	\label{eq:n_cl}
\end{equation}
where $\rho_\text{loc}=0.3$~GeV/cm$^3$.

The analogous results for some other models are given in tab.~\ref{tab}. For all the models ``core''-radius and clump size were taken the same as in BGZ model. 
It is seen that the BGZ model predicts the minimal number of PGSs. So other models become more sensitive to observational data and the comparison with the data becomes very important tool for probing them.

\begin{figure*}[t]
  \centering
  \subfigure[The case without y-interaction is shown. \label{fig:beta1}]{\includegraphics[width=0.48\textwidth]{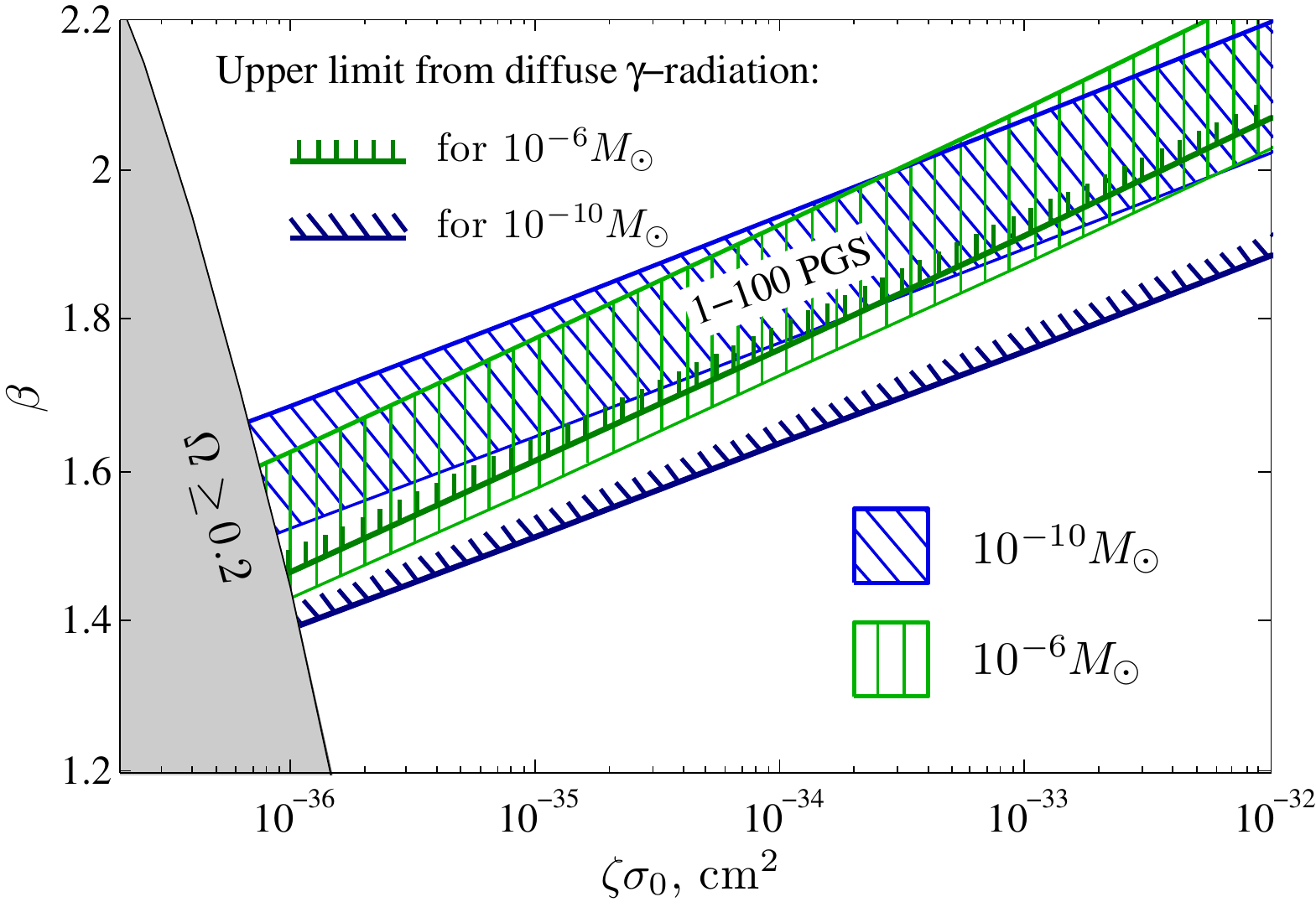}}\quad
  \subfigure[The case with y-interaction is shown. \label{fig:beta2}]{\includegraphics[width=0.48\textwidth]{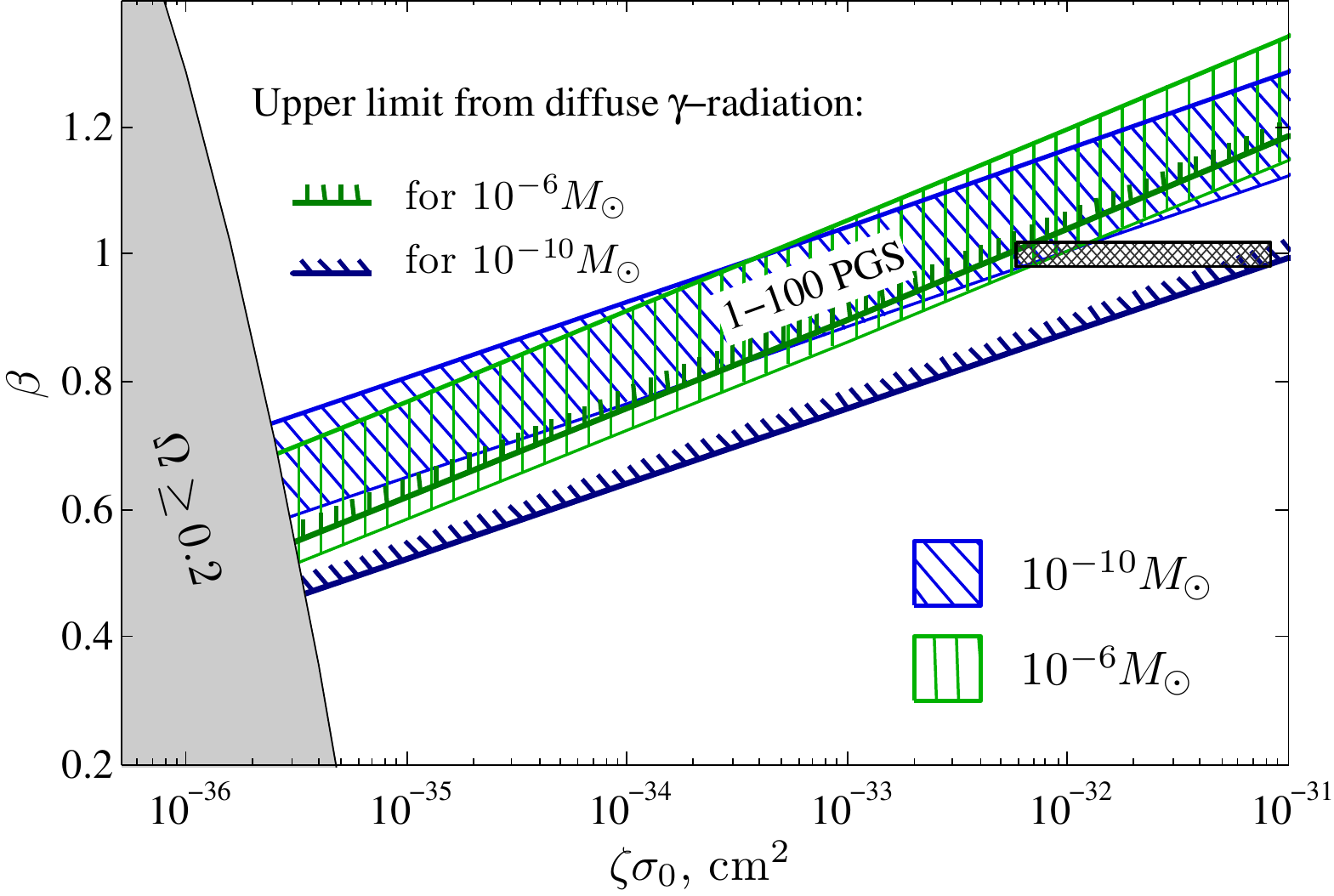}}
  \caption{Allowed and forbidden regions of parameters $\beta$ and $\sigma_0$ are shown as they obtained on the base of Fermi LAT data on PGS and diffuse $\gamma$-radiation. Two values of the minimal clump mass are given. $\nu_4$ on fig.~\ref{fig:beta2} corresponds to the cases of heavy neutrino in the mass range 45--49~GeV.
  }
  \label{fig:beta}
\end{figure*}

The spatial distribution of clumps at distance $l_\text{max}$ from the Earth is expected to be homogeneous, therefore its distribution on celestial sphere should be isotropic. Distribution of unidentified PGS registered by Fermi LAT \cite{2012ApJS..199...31N} is almost isotropic except of region of the galactic plane. The isotropic component include $\sim$100 sources. Supposing that the nature of all of them can be related to dark matter clumps, it is possible to determine regions of the parameters magnitudes $\beta$ and $\sigma_0$ for the typical clump masses $10^{-10} \div 10^{-6} M_{\odot}$ \cite{2005Natur.433..389D}. The results are shown at fig.~\ref{fig:beta} where the factor $\zeta$ is
\begin{multline}
	\zeta = \left( \frac{m}{100\ \text{GeV}} \right)^2\left(\frac{\Omega_{\text{CDM}}}{0.2}\right)^2
	\left(\frac{10}{N_{\gamma}}\right)\times\\
	\times\left(\frac{F_{\text{min}}}{3\cdot 10^{-9}\ \cm^{-2} \s^{-1}}\right)
	\left( \frac{0.002}{\xi} \right)^{2/3}.
\end{multline}
This factor includes uncertainties of the chosen parameters of DM particle properties (not of clump density profile). The predicted number of the clumps showing themselves as PGS depends on it as $N\propto\zeta^{-2/3}$. It is useful to take it into account while considering results for other models (see tab.~\ref{tab}).

The clumps which are not visible (farther than $l_\text{max}$) should contribute in diffuse $\gamma$-radiation. $\gamma$-Flux from them in given solid angle can be expressed as
\begin{equation}
	\Phi= \int^{l_{\rm halo}}_{l_\text{max}} F n_{\text{cl}} l^2 dl = \frac{P n_{\text{cl}}l_{\rm halo}^{\rm eff}}{4\pi},
	\label{eq:Phi}
\end{equation}
where $F$ and $P$ are introduced in Eq.~\ref{eq:flux}, $l_{\rm halo}$ is the distance to the edge of halo along to line of sight and $l_{\rm halo}^{\rm eff}\approx 10$ kps is its effective value (typical for many halo density profiles); $l_\text{max}$ is negligible with respect to $l_{\rm halo}$.
One requires that
\begin{equation}
    \Phi<\Phi_{\rm exp}\approx 1.5\cdot 10^{-5}\ \text{cm}^{-2}\text{s}^{-1}\text{sr}^{-1},
    \label{eq:Phi_exp}
\end{equation}
where $\Phi_{\rm exp}$ is the diffuse $\gamma$-background measured by LAT \cite{2009PhRvL.103y1101A}.

Eq.~\ref{eq:Phi_exp} gives upper limit in the plot fig.~\ref{fig:beta}. As seen, the case of clump mass $M=10^{-10}M_{\odot}$ is fully excluded, the case of $M=10^{-6}M_{\odot}$ is constrained but up to $\sim$10 PGS are still possible. Higher masses avoid such constraining.

As to the parameters $\sigma_0$ and $\beta$, the range $3\cdot 10^{-34}\lesssim\sigma_0\lesssim 10^{-32}$~cm$^2$ and $\beta=2$ ($\beta=1$) in case without (with) y-interaction are found to be the most interesting.

\begin{figure}[t]%
	\centering
	\includegraphics[width=0.5\textwidth]{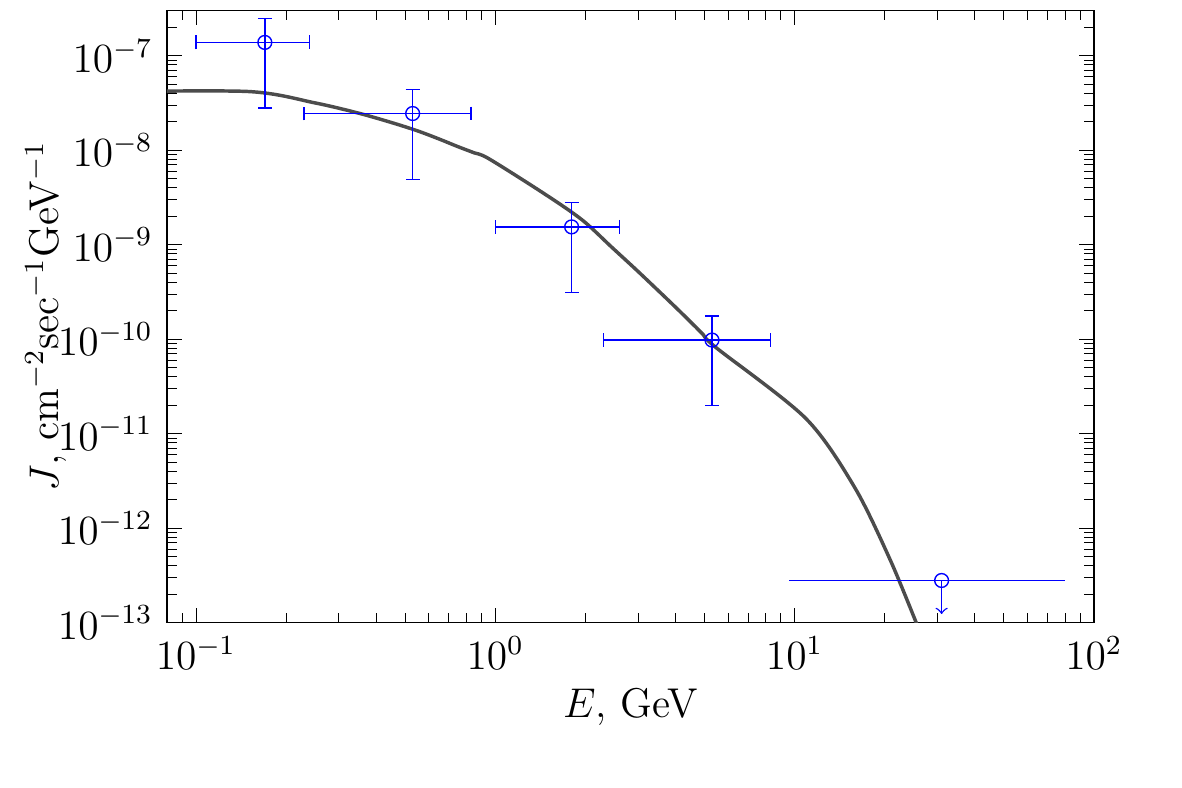}%
	\caption{Expected spectrum from $\nu_4\bar\nu_4$-annihilation
    is shown in comparison with the observed spectrum of nonidentified source 2FGL J1653.6-0159.}%
	\label{fig:NeutrinosSpectrum}%
\end{figure}

Several specific models of annihilating DM particles have been considered: neutralino \cite{1996PhR...267..195J}, heavy neutrino \cite{2008PAN....71..147B, 2002PAN...65..382} (with y-interaction and without it), Kaluza-Klein particles \cite{2003NuPhB.650..391S}, dark atoms OHe \cite{2006GrCo...12...93B, 2006PZETF...83..3}.
All candidates except for a heavy neutrino ($\nu_4$) with additional interaction cannot explain LAT data but, at the same time, do not contradict them in the case of BGZ profile. The case of $\nu_4$ is pointed on fig.~\ref{fig:beta2}. As seen, the part of nonidentified point-like LAT sources may be explained due to $\nu_4\bar{\nu}_4$-annihilation at mass $m_{\nu_4}\sim 46-49$~GeV. Typical $\gamma$-spectrum from 47 GeV neutrinos annihilation is shown at the fig.~\ref{fig:NeutrinosSpectrum} in comparison with measured spectrum of one of the non-identified PGS (annihilation spectrum was obtained with the help of Monte-Carlo generator Pythia 6.4 \cite{url:Pythia}).
It is worth to note that the heavy neutrino parameters are strongly restricted by underground experiments \cite{2008PAN....71..147B, 2009PhRvL.102a1301A}. At the same time the prediction of $\nu_4$ relic density suffers with uncertainty connected with recombination of the pair of y-interacting neutrino-antineutrino in early Universe \cite{2008PAN....71..147B}.

\section{Some possible features of the model}
\subsection{Spatially extended sources}

The distribution of clumps does not exclude a possibility of their presence in close vicinity to Solar system. Therefore an angular size of such the sources can exceed the angular resolution of detector $\delta$ (corresponding to the solid angle, say, $\Omega_{\delta}=\pi (\delta/2)^2$).

In the capacity of criterion of extended gamma-source (EGS) we choose the following. The minimal amount of photons needed to recognize a source over background from a region $\Omega_{\delta}$ is defined by the minimal point source flux $F_\text{min}$. Statistically significant signal from region with the size of $\nu\Omega_{\delta}$ must be then at least $>\sqrt{\nu} F_{\rm min}$.
For clump candidate to EGS we require fulfilment of similar condition for the circle region of $\pi\delta^2=4\Omega_{\delta}$ size around the clump center
excluding the $\pi (\delta/2)^2$ center part:
\beq
  F(4\Omega_{\delta})-F(\Omega_{\delta})\geq 2 F_\text{min}.
  \label{eq:NptCrit}
\eeq
Here for minimal flux a little harder condition was taken than it should be proceeding from the number of $\Omega_{\delta}$ regions which the ring $[\delta/2\ldots \delta]$ covers, i.e. $\sqrt{3}F_\text{min}$. However, this criterion (\ref{eq:NptCrit}) is softer than requirement $F(\Omega_{\delta})\geq F_\text{min}$ for several adjacent $\Omega_{\delta}$ regions simultaneously or even $F(\nu\Omega_{\delta})\geq \nu F_\text{min}$ for all the $\nu\Omega_{\delta}$ region wholly, and it needs a special analysis of observation data.

\begin{figure}[t]
    \centering
    \includegraphics[width=0.48\textwidth]{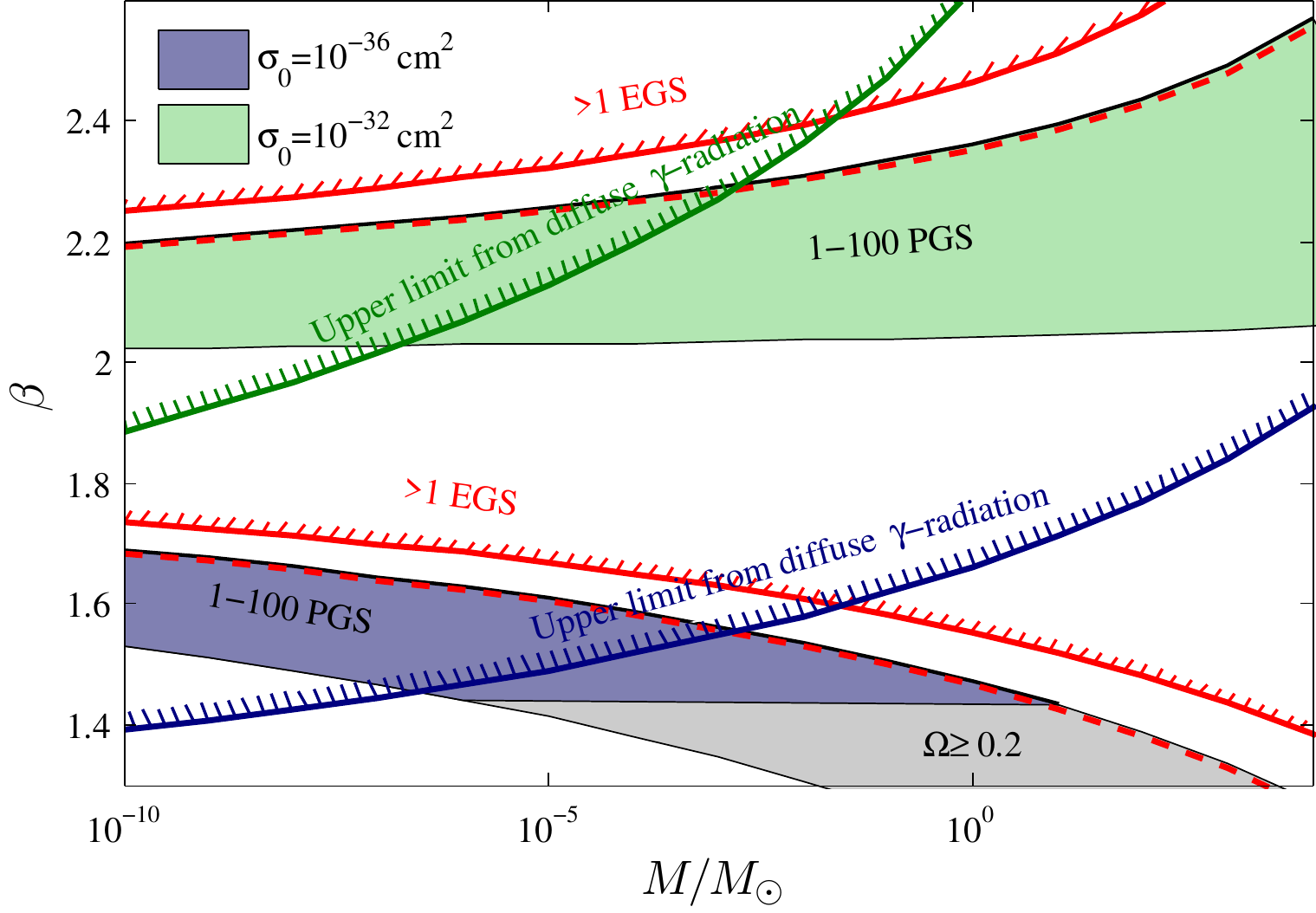}
    \caption{Allowed and forbidden regions of parameters $\beta$ and $M$ are shown for two typical values of $\sigma_0$ without y-interaction. The solid and dashed near red lines show 1 extended source for Fermi LAT and Gamma-400 detectors respectively.}
    \label{fig:BetaM}
\end{figure}

Based on Eq.~\eqref{eq:NptCrit} we find that Fermi LAT can hardly observe any clump as non-point-like source, what does agree with observation data. However, it becomes possible at the improved in future experiment values of $F_\text{min}$ and $\delta$. At the fig.~\ref{fig:BetaM} we show the regions of parameters of clump mass $M$ and $\beta$ satisfying to data on diffuse and discrete sources of $\gamma$-radiation. Dashed line shows EGS discovery potential at the achievement of $\delta=\delta_\text{G400}\approx 0.3^\circ$ (vs $\delta_\text{LAT}\approx 0.6^\circ$ for Fermi LAT at $E_\gamma\gtrsim 1$~GeV \cite{2009ApJ...697.1071A}) and the same $F_\text{min}$ as for Fermi LAT. In principle, such values can be reached for Gamma-400 detector \cite{2012arXiv1201.2490G} at some energy interval. EGS discovery will help to test the clump model.

Note, that improvement (minimization) of $F_\text{min}$ (in perfect case, minimal flux per unit of solid angle) is found to be more promising from viewpoint of EGS discovery than improvement of $\delta$ only. However, at better $\delta$ one can partially compensate low sensitivity to photon flux by taking several $\Omega_{\delta}$ region.

\subsection{``Traveling'' sources}

Provided DM clump accounts for PGS, there can be sources which change visibly its position on celestial sphere within time of observations (the time between EGRET experiment \cite{1999ApJS..123...79H} and Fermi LAT is $\sim$10~years), due to its closeness to Solar system. Clump velocity in Galaxy is $v\sim 250$~km/s. For 10~years they travel the distance about $\Delta\sim 2.6\cdot 10^{-3}$~pc. Let us estimate the number of clumps with visible movement. The number of clumps per the spherical layer $r\div r+dr$ and per interval $\cos\theta\div\cos\theta+d\cos\theta$, where $\theta$ is the angle between radius-vector $\vec{r}$ (originating from observer) and vector of clump velocity $\vec v$, is
\beq
    \frac{dN}{dr d\cos\theta} = 4\pi r^2 n_{\rm cl}\frac{1}{2}.
    \label{eq:d2N}
\eeq
Here we refer to angular distribution of $\vec{v}$ as isotropic one with $\theta$ ranging from 0 to $\pi$.

Integration \eqref{eq:d2N} gives the distribution in angle
$$
  \phi=\phi(r,\theta)\approx \frac{\Delta\sin\theta}{r},
$$
which clump travels on the sky,
\begin{multline}
	\frac{dN}{d\phi} = \int\frac{dN}{dr d\cos\theta}drd\cos\theta\delta\left(\phi-\phi(r,\theta)\right)=\\
	= \frac 34\pi^2\frac{\Delta^3}{\phi^4}n_{\rm cl}.
\end{multline}
Here we changed the variables $r$ and $\cos\theta$ to $\phi$ through integration with $\delta$-function (divergence at $\phi \rightarrow 0$ is a consequence of our approximation $n_{\rm cl}= \textrm{const}$ for $0<r<\infty$).

Estimation of the number of the moved (visibly) sources gives
\begin{multline}
	N_\text{trav} = \int\limits_{\phi_{\rm min}}^{\phi_{\rm max}} \frac{dN}{d\phi}d\phi \approx \frac 14\pi^2  n_{\rm cl} \frac{\Delta^3}{\phi^3_{\rm min}}\le\\
	\le 0.4\left(\frac{10^{-7}M_\odot}{M}\right).
	\label{eq:NShift}
\end{multline}
Here $\phi_{\rm max}\gg 1$ does not affect the result, $\phi_{\rm min} = \max(\delta_{\text{resol}},\delta_{\text{lim}} )$, where $\delta_{\text{resol}}=\sqrt {\delta_{\rm EGRET}^2 +\delta_{\rm LAT}^2}\sim1.47^\circ$ with $\delta_{\rm EGRET}\approx1.34^\circ$ and $\delta_{\rm LAT}\approx0.6^\circ$ being the angular resolutions of detectors EGRET and LAT for $E_{\gamma}\geqslant1$ ~GeV, $\delta_{\text{lim}}\sim \Delta/l_{\max}$ corresponds to angular shift of the clump at maximal observable distance. For maximal estimations in \eqref{eq:NShift} we put $\phi_{\rm min} = \delta_{\text{resol}}$, that gave result independent on the model of the clump density profile. In case of $\phi_\text{min}=\delta_{\text{lim}}>\delta_{\text{resol}}$, all the visible clumps (if they are) should be traveling.

Thus, as seen from \eqref{eq:NShift}, for mass of clumps $M \sim 10^{-7} M_\odot$ the probability to find the moved sources becomes noticeable (for any density profile). The value $M \sim 10^{-7} M_\odot$ is close to the minimal one which is allowed from data on diffuse $\gamma$-background (fig.~\ref{fig:BetaM}). However, time of observation(s) grows and future experiment like Gamma-400 will work at better angular resolution. Both from the time and from the resolution the expected number of traveling sources (TS) depends as a cube power (see \eqref{eq:NShift}). It challenges for future experiment to probe the model with searching for TS for clump mass up to $\sim (10^{-6}\div 10^{-5}) M_\odot$. Even now, one may consider time interval of 20 years (from the start of EGRET) and expect to find several TS.

In case of $N_\text{trav}\gg 1$ the movement of the sources should have a regular character accounted for by solar system motion around Galactic center (GC). Respective data analysis would provide much more strong test for discussed model, since virtually all the known astrophysical explanations of PGSs do not relate them with objects of halo (but rather with galactic disc or distant galaxies, in case of which observations are insensitive to solar system motion around GC).

Existing difference in data on non-identified PGSs of EGRET and LAT (LAT confirms existence of only 30--40\% of the sources from catalogue 3EG \cite{1999ApJS..123...79H, 2012ApJS..199...31N} and observes new sources which had not been registered by EGRET) could be explained by effects of traveling clump-sources in small part.

\section{Conclusion}

In this paper we refine the previous results \cite{2013YadPhys...76...286} and have shown that DM clumps could be point-like (and extended) sources of the $\gamma$-radiation and they can partially explain non-identified $\gamma$-sources, registered by LAT and EGRET. The proposed method allows to estimate an amount of accessible to observation of DM clumps for various models of DM particle and density profile.
Note that the suppression of the subdominant fraction of DM particles in clumps of mass $M<M_{\rm min}$ (if they are) has not been taken into account and requires special research.

The values of parameters $\sigma_0$ and $\beta$, at which DM model is either consistent or inconsistent with the LAT data, have been determined for the most ``conservative'' model BGZ \cite{2003PhRvD..68j3003B}. The high sensitivity of the predictions to the choice of density profile model is shown.

The clumps, situated in a close vicinity of Solar system, may account partially for noncoincidences between catalogues EGRET and LAT (the sources, registered by EGRET and not confirmed by LAT) and also for a spatially extended gamma-ray sources which can be detected by future gamma-telescopes.

Possibilities that gamma-radiating clump changes visibly its position on celestial sphere and it is seen as a spatially extended gamma-source
can play a role of distinctive features, to be probed in future experiments. It would allow to distinguish from alternative models of possible gamma-source origin and in particulary the model associated with primordial black hole clusters \cite{2011GrCo...17...27B,2011APh....35...28B}.

\section{Acknowledgements}

The authors express gratitude to V.~Dokuchaev and Yu.~Eroshenko for useful discussion of clump evolution. The work of A.K. and K.B. was supported by grant of RFBR \textnumero~12-02-12123 and by The Ministry of education and science of Russia, projects \textnumero~14.132.21.1446, \textnumero~8525 and \textnumero~14.A18.21.0789.

\bibliographystyle{unsrt}
\bibliography{Article}

\end{document}